\documentclass[%
 reprint,
 amsmath,amssymb,
 aps,
prb,
superscriptaddress
]{revtex4-2}

\usepackage[utf8]{inputenc} 
\usepackage[T1]{fontenc}    

\usepackage{graphicx}
\usepackage{dcolumn}
\usepackage{bm}
\usepackage{hyperref}
\usepackage{amsmath}
\usepackage{siunitx}
\usepackage{tabularx}
\usepackage{multirow}

\begin{document}
    \newcommand{\YbZnGaO}{YbZn$_{2}$GaO$_{5}$}
    \newcommand{\LuZnGaO}{LuZn$_{2}$GaO$_{5}$}
    \newcommand{\Yb}{Yb$^{3+}$}
    \newcommand{\oxygen}{O$^{2-}$}
    
    \title{Quenched Disorder in the Triangular Lattice Antiferromagnet \YbZnGaO}

    \author{Leshan Zhao}
    \email{lzhao53@jhu.edu}
    \author{Tong Chen}
    \affiliation{William H. Miller III Department of Physics \& Astronomy, Johns Hopkins University, Baltimore, Maryland 21218, USA}
    \author{Matthew B. Stone}
    \author{Qiang Zhang}
    \author{Colin L. Sarkis}
    \affiliation{Neutron Scattering Division, Oak Ridge National Laboratory, Oak Ridge, Tennessee 37831, USA}
    \author{S. M. Koohpayeh}
    \affiliation{William H. Miller III Department of Physics \& Astronomy, Johns Hopkins University, Baltimore, Maryland 21218, USA}
    \affiliation{Department of Materials Science and Engineering, Johns Hopkins University, Baltimore, Maryland 21218, USA}
    \affiliation{Ralph O’Connor Sustainable Energy Institute, Johns Hopkins University, Baltimore, Maryland 21218, USA}
    \author{Collin Broholm}
    \email{broholm@jhu.edu}
    \affiliation{William H. Miller III Department of Physics \& Astronomy, Johns Hopkins University, Baltimore, Maryland 21218, USA}

    \date{\today}

    \begin{abstract}
        We investigate the crystal electric field (CEF) excitations of \Yb\ ions in powder samples of the triangular-lattice rare-earth-based antiferromagnet \YbZnGaO\ using inelastic neutron scattering (INS). Three CEF excitations from the ground-state Kramers doublet were observed, each exhibiting significant broadening beyond instrumental resolution. Combining temperature-dependent INS and neutron powder diffraction, we identify a significant static contribution to this broadening and attribute it to heterogeneous coordination of \Yb\ ions due to Ga$^{3+}$/Zn$^{2+}$ site mixing. Rietveld refinement of neutron powder diffraction indicates that 35\% of Ga occupies the Zn site and 60\% of Zn occupies the Ga site. We show with a point charge model for the CEF Hamiltonian that heterogeneous coordination of \Yb\ ions leads to broadened CEF peaks. First-principles calculations demonstrate that the random Ga$^{3+}$/Zn$^{2+}$ distribution can produce the distortions of the YbO$_6$ octahedra observed from neutron diffraction. Because the documented heterogeneity will extend to exchange interactions, our results suggest that disorder is a significant factor in the unusual magnetism previously reported in \YbZnGaO, including broad low-energy magnetic excitations and the absence of magnetic ordering down to \SI{0.3}{K}.
    \end{abstract}

    \maketitle

    \section{Introduction}

   Quantum spin liquids (QSLs) are an exotic low temperature state of matter where symmetry-breaking long-range magnetic order may be supplanted by long-range quantum entanglement and magnons can fractionalize into spin-1/2 spinons\cite{doi:10.1126/science.aay0668}.  The search for QSLs has focused on materials where magnetic interactions are frustrated due to lattice geometry as in the triangular and kagome lattices, where competing interactions can lead to a highly degenerate manifold of low energy states\cite{Balents2010}. Although many candidate materials have been reported, magnetic or nonmagnetic defects, spatial anisotropy, or interlayer magnetic interactions often complicate interpretation of the experimental results. In particular, it can be difficult to distinguish static disordered forms of magnetism from intrinsic long-range quantum entangled states of matter\cite{Lee2007,Li2012,Li2013,Freedman2010,Gomilek2016,Li2014,Shimizu2003,Fortune2009,Doi2004}. Therefore, in the ongoing quest to realize this new state of matter, it is important to characterize and understand quenched disorder and its effects on collective phenomena in QSL candidate materials.\cite{doi:10.1021/acs.chemrev.0c00641} 

    \YbZnGaO\ contains a quasi-two-dimensional triangular lattice of \Yb\ ions with effective spin-1/2 moments and is an interesting new QSL candidate\cite{PhysRevLett.133.266703}. Magnetic susceptibility measurements show no signs of static magnetic ordering down to \SI{0.3}{K}. Specific heat measurements display a quadratic temperature dependence at low temperatures with no residual entropy down to \SI{0.06}{K}  and inelastic neutron scattering (INS) measurements on single crystalline samples show a gapless continuum of scattering at low energies, consistent with fractionalized excitations\cite{PhysRevLett.133.266703}. These results are broadly consistent with theoretical predictions for a U(1) Dirac QSL.\cite{PhysRevB.93.144411,PhysRevB.108.L220401,PhysRevX.14.021010}

    However, INS measurements of the higher-energy crystal electric field (CEF) excitations in \YbZnGaO\ show significant physical broadening of the peaks beyond instrumental resolution\cite{PhysRevLett.133.266703}. Similar broadening has been reported in related quasi-two-dimensional triangular lattice compounds based on Yb, such as YbZnGaO$_{4}$ and YbMgGaO$_{4}$, where it has been attributed to charge disorder near \Yb\ ions stemming from non-magnetic site mixing \cite{PhysRevLett.118.107202, paddison_daum_dun_ehlers_liu_stone_zhou_mourigal_2016,PhysRevLett.120.087201}. A theoretical study has shown that such disorder may produce a heterogeneous disordered state without long-range entanglement or magnon fractionalization yet with inelastic magnetic neutron scattering similar to that of a QSL\cite{PhysRevLett.119.157201}. 

    In this work, we report an investigation of the CEF excitations and the crystal structure of \YbZnGaO\ to understand the role of disorder in its unusual magnetic properties. We identify a strong static contribution to the broadening of the CEF peaks and link it to a disordered local charge environment of \Yb\ arising from $\rm Ga^{3+}/Zn^{2+}$ site mixing. We support this model by neutron powder diffraction and first-principle calculations. These findings provide crucial context for interpreting the QSL-like experimental properties previously reported for this material.

    \section{Methods}

    \subsection{Sample preparation}

\begin{figure}
        \centering
        \includegraphics[width=0.9\linewidth]{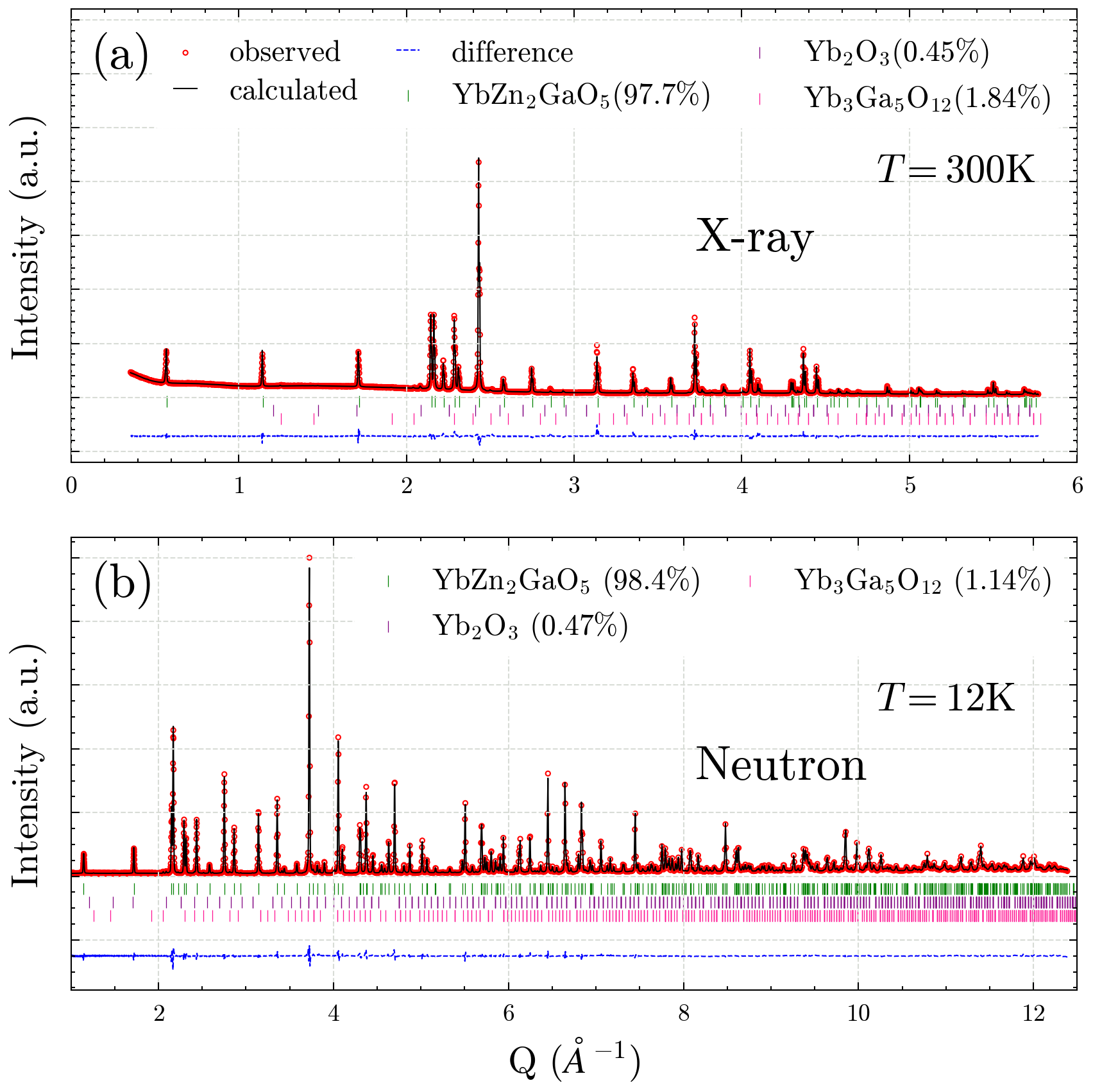}
        \caption{(a) X-ray powder diffraction patterns of \YbZnGaO\ acquired at room temperature refined with the Rietveld method. Vertical bars mark the location of Bragg peaks for the majority and two minority phases. The blue dashed line is the difference between measurement and refinement. A phase purity of \YbZnGaO\ of \SI{97.7}{wt\%} along with small impurities of Yb$_2$O$_3$ (\SI{0.45}{wt\%}) and Yb$_3$Ga$_5$O$_{12}$ (\SI{1.84}{wt\%}) were observed. Lattice constants of $a=$\SI{3.376}{\angstrom} and $c=$\SI{21.96}{\angstrom} were obtained for \YbZnGaO. (b) Rietveld-refined neutron powder diffraction pattern of \YbZnGaO\ measured at $T=$ \SI{12}{K}. The majority phase \YbZnGaO\ constitutes \SI{98.4}{wt\%} of the sample. The impurity phases are Yb$_2$O$_3$ (\SI{0.47}{wt\%}) and Yb$_3$Ga$_5$O$_{12}$ (\SI{1.14}{wt\%}). The unit cell model used in the refinement allows for Zn and Ga atoms to partially occupy each other's Wyckoff sites. The best fit was achieved with Zn  occupying 60(1)\% of the Ga site, and Ga occupying 35(1)\% of the Zn site. The refinement parameters obtained from the low $T$ neutron data are in Table \ref{tab:neutron_diffraction}.}
        \label{fig:diffraction}
    \end{figure}

    Polycrystalline \YbZnGaO\ was synthesized via a solid-state reaction method. Stoichiometric amounts of high-purity precusors Yb$_{2}$O$_{3}$(99.9\%), ZnO(99.9\%), and Ga$_{2}$O$_{3}$(99.9\%) were thoroughly mixed and pressed into a pellet. The pellet was sintered at \SI{1225}{\celsius} for a total of 213 hours with intermediate grinding. Phase purity was confirmed with powder X-ray diffraction, as shown in Fig.~\ref{fig:diffraction}(a). The sample is primarily single phase with minor impurities of $\rm Yb_2O_3$ (\SI{0.45}{wt\%}) and $\rm Yb_3Ga_5O_{12}$ (\SI{1.84}{wt\%}).

   \begin{figure}[t]
        \centering
        \includegraphics[width=0.9\linewidth]{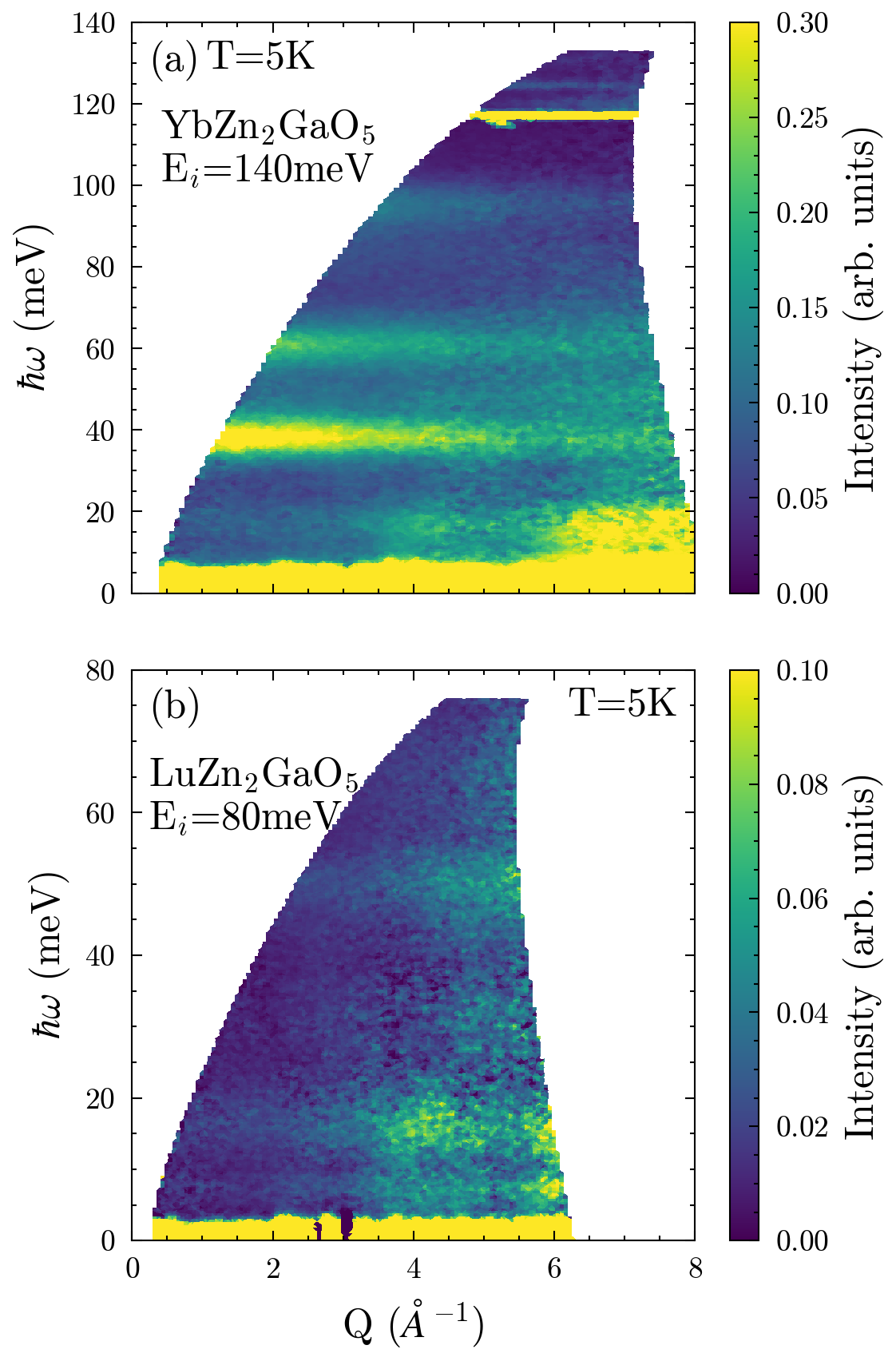}
        \caption{(a) Inelastic neutron scattering spectrum of \YbZnGaO\ measured at $T=$ \SI{5}{K} with incident neutron energy $E_i=$ \SI{140}{meV}, revealing three crystal field excitations at $\hbar\omega=$ \SI{38}{meV}, \SI{61}{meV}, and \SI{94}{meV}. The line at \SI{118}{meV} is not associated with the sample but with a fast neutron pulse from the proton accelerator complex. (b) Inelastic neutron scattering spectrum of \LuZnGaO\ measured at $T=$ \SI{5}{K} with incident neutron energy $E_i=$ \SI{80}{meV}, used to subtract phononic contribution from the \YbZnGaO\ data. Note that a lower $E_i$ was used for \LuZnGaO\ to avoid a \SI{142}{meV} resonant neutron absorption process in Lu nuclei.}
        \label{fig:INS_spectra}
    \end{figure}

    \subsection{Neutron powder diffraction}
    Neutron powder diffraction measurements were conducted using the neutron diffractometer POWGEN at the Spallation Neutron Source (SNS) at Oak Ridge National Laboratory (ORNL)\cite{Huq2019}. An automatic sample changer (PAC) was used as the sample enviornment to cover a temperature range between \SI{12}{K} and \SI{300}{K}. Neutron bank 2 with a center wavelength of \SI{1.5}{\angstrom } was used
    to collect data at temperatures of $T=$ \SI{12}{K}, \SI{50}{K}, \SI{100}{K}, \SI{150}{K}, and \SI{200}{K}. Structural refinements were performed using the Rietveld method as implemented in the FullProf suite\cite{RodrguezCarvajal1993}.

    \begin{figure}[t]
        \centering
        \includegraphics[width=0.9\linewidth]{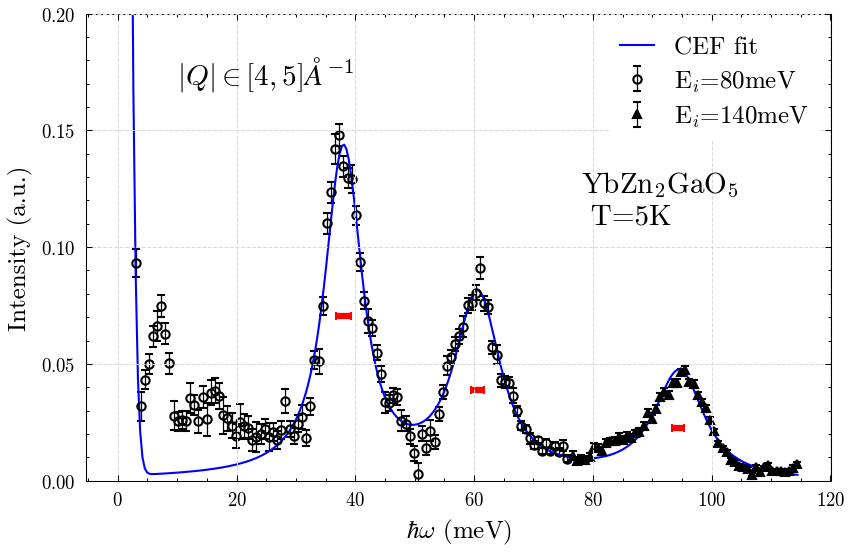}
        \caption{Background- and phonon-subtracted inelastic neutron scattering spectrum of a \YbZnGaO\ powder sample acquired at $T=$ \SI{5}{K}. A $Q-$ and $\hbar\omega-$dependent absorption correction was applied to each data set and the data were averaged over $Q\in[4,5]$~\AA$^{-1}$. The blue line represents a fit based on a crystal field Hamiltonian with each delta-function replaced by a Voigt peak with the calculated instrumental resolution fixed as its Gaussian width. The fitted crystal field parameters are listed in Table \ref{tab:Bmn}. The fitted peaks show peak widths (FWHMs) significantly beyond the calculated instrumental resolution, which is indicated by red horizontal bars. The weak peaks for $\hbar\omega <20$~meV are residual Yb/Lu phonon scattering associated with the three times larger nuclear scattering cross section for Yb as compared to Lu. The intrinsic broadening of the CEF excitations was extracted as the Lorentzian FWHM $\Delta E_{L}$ of the Voigt profiles, and is listed in Table \ref{tab:CEF_FWHM}.}
        \label{fig:INS_Ecut}
    \end{figure}

    \begin{figure}[t]
        \centering
        \includegraphics[width=0.9\linewidth]{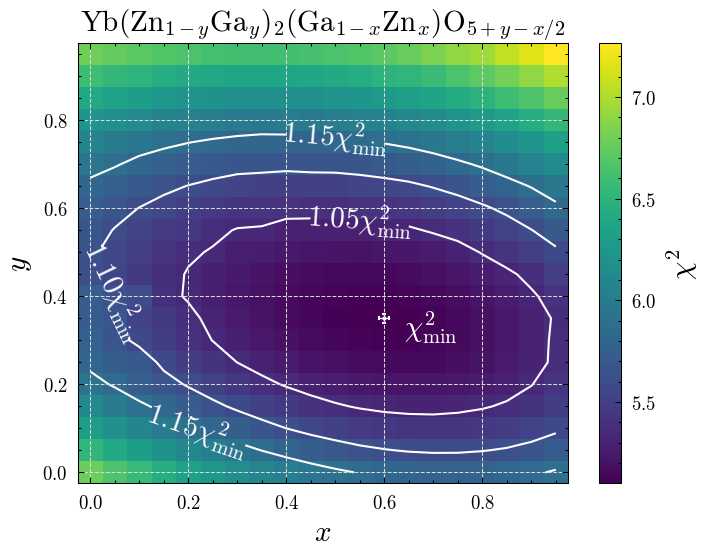}
        \caption{$\chi^2(x,y)$ goodness of fit for the Rietveld refinement of the $T=12$~K neutron powder diffraction data in Fig.~\ref{fig:diffraction}(b) versus site mixing between Zn and Ga in $\rm Yb(Zn_{1-y}Ga_y)_2(Ga_{1-x}Zn_x)O_{5+y-x/2}$. All other structural parameters were optimized at each point in the $(x,y)$ grid. The small white dot with error bars indicates the minimum of $\chi^2$, denoted as $\chi^2_{\rm min}$, which is found where Zn atoms occupy $x=60(1)$\% of the Ga site, and Ga atoms occupy $y=35(1)$\% of the Zn site. White contour lines denote where  $\chi^2(x,y) = 1.05\chi^2_{\rm min}$, $1.1\chi^2_{\rm min}$, and $1.15\chi^2_{\rm min}$. Notably, the structure model with no $\rm Ga^{3+}/Zn^{2+}$ site mixing, corresponding to $(x,y)=(0,0)$,  yields a $\chi^2$ that is more than 30\% greater than $\chi^2_{\rm min}$.}
        \label{fig:neutron_diffraction_chi2}
    \end{figure}

    \subsection{Inelastic neutron scattering}
    \label{sec:INS}
    Inelastic neutron scattering measurements were carried out using the Fine-Resolution Fermi Chopper Spectrometer (SEQUOIA)\cite{Granroth_2010} at SNS, ORNL. \SI{6}{g} each of \YbZnGaO\ and its isostructral, nonmagnetic analogue \LuZnGaO\ were sealed at room temperature in aluminum sample cans with an atmosphere of helium gas and mounted along with an empty sample can in a bottom-loading closed-cycle refrigerator with a three-sample changer. Measurements were performed using incident neutron energies of $E_{i}=$ \SI{80}{meV} and \SI{140}{meV} for \YbZnGaO, and $E_{i}=$ \SI{80}{meV} for \LuZnGaO\ over a temperature range of $T\in [5,300]$~K in intervals of \SI{60}{K}. For $E_{i}=$ \SI{80}{meV}, the high resolution Fermi chopper was used with a frequency of $\nu=$\SI{480}{Hz}, producing full width at half maximum (FWHM) elastic energy resolution of $\Delta E_{\rm res}=$ \SI{3.1}{meV}. For $E_{i}=$ \SI{140}{meV}, the high flux Fermi chopper was used with a frequency of $\nu=\SI{480}{Hz}$, resulting in $\Delta E_{\rm res}= \SI{4.7}{meV}$ at the elastic line.  Contributions from the sample can were subtracted using data collected from the empty sample can. The data were corrected for sample absorption with a Monte Carlo method as implemented in MANTID\cite{ARNOLD2014156}. For part of the \YbZnGaO\ spectrum with energy transfer $\hbar \omega <\SI{80}{meV}$, phonon contributions were subtracted using \LuZnGaO\ data. Because the nuclear scattering cross section for Yb ($\sigma({\rm Yb})=23.4$~b) is much larger than for Lu ($\sigma({\rm Lu})=7.2$~b), phonons at low $\hbar\omega$ that involve motion of Yb/Lu produce more neutron scattering for \YbZnGaO\ than for \LuZnGaO\ and therefore will not be removed in the background subtraction process. This is apparent in Fig.~\ref{fig:INS_Ecut} for $\hbar\omega<20$~meV. For $\hbar\omega>\SI{80}{meV}$, where \LuZnGaO\ data were not available due to practical limits on $E_i$ set by a \SI{142}{meV} resonant neutron absorption process in lutetium, a linearly interpolated background was subtracted. Finally, data acquired with different incident energy were normalized to each other using the integrated intensity of the two lowest-energy crystal field excitations.

    \subsection{First-principle calculations}
    First-principle calculations were performed using the QUANTUM ESPRESSO package\cite{Giannozzi_2009} with Perdew-Burke-Ernzerhof (PBE) type exchange-correlation potentials. A projector augmented wave (PAW) pseudopotential optimized for rare-earth elements was used for Yb to describe its localized $4f$ electrons\cite{TOPSAKAL2014263}. The plane-wave cutoff energy was set to \SI{120}{Ry} and the Brillouin zone was sampled with a $6\times 6\times 2$ k-point mesh. The lattice parameters and atomic positions were fully relaxed until the forces on all atoms were less than \SI{e-5}{Ry/\angstrom }. The pseudo-disordered structures considered in the calculations were periodic, charge neutral, and preserved both the stoichiometry and the three-fold rotational symmetry of the crystal structure.
    \label{subsec:DFT}

    \section{Experimental results}

    \subsection{Broadened crystal field transitions}

    As captured by Hund's rules, a free \Yb\ ion has a spin angular momentum of $S=1/2$ and an orbital angular momentum of $L=3$. Spin-orbit interactions yield an   eight-fold degenerate ($J=7/2$) spin-orbital ground state. In the nominal structure of \YbZnGaO, \Yb\ occupies a single Wyckoff site with a trigonal $D_{3d}$ point group symmetry. The corresponding crystal electric field (CEF) should split the eight-fold degenerate $J=7/2$ multiplet into four Kramers doublets, resulting in three dipolar-active excitations from the ground state that should be visible as sharp peaks in the inelastic magnetic neutron scattering spectrum.

    Sample can-subtracted INS spectra of \YbZnGaO\ and its isostructral, nonmagnetic analogue \LuZnGaO\ collected at $T=$ \SI{5}{K} are shown in Fig.~\ref{fig:INS_spectra}. Three distinct CEF excitations are indeed observed for \YbZnGaO\ but not for non-magnetic \LuZnGaO\ at $ E_{\rm exp}=$ \SI{38.0(1)}{meV}, \SI{60.5(2)}{meV}, and \SI{94.3(2)}{meV}, consistent with previous reports\cite{PhysRevLett.133.266703}. After the background, absorption, and normalization corrections described in Section ~\ref{sec:INS}, the magnetic excitation spectrum for \YbZnGaO\ averaged over momentum transfer $Q \in [4,5]$ \SI{}{\angstrom^{-1}} is shown in Fig.~\ref{fig:INS_Ecut}. 
    
    All three CEF excitations display significant broadening, with FWHM $\Delta E_{\rm exp}=$ \SI{7.5(3)}{meV}, \SI{9.4(5)}{meV}, and \SI{10.1(4)}{meV}, respectively, as obtained from a Voigt fit to the peaks. These are all well beyond the corresponding instrumental resolution of $\Delta E_{\rm res}=$ \SI{2.4}{meV}, \SI{2.2}{meV}, and \SI{2.0}{meV}, which are shown by red bars in Fig.~\ref{fig:INS_Ecut}. Such broadening is not anticipated in a pure crystalline dielectric solid at low temperatures.

\subsection{$\rm Ga^{3+}/Zn^{2+}$ site mixing}
   To determine the origin of this broadening,  we looked for indications of structural disorder using neutron powder diffraction. Zn and Ga are immediate neighbors in the periodic table so conventional x-ray diffraction provides little contrast. Neutrons, however, can distinguish Zn from Ga due to the different bound coherent neutron scattering lengths for their nuclei: $b_{\rm Ga}=\SI{7.288}{fm}$ while $b_{\rm Zn}=\SI{5.680}{fm}$. The corresponding measure of contrast for neutrons $2(b_{\rm Zn}-b_{\rm Ga})/(b_{\rm Zn}+b_{\rm Ga})=25$\% thus exceeds that for x-rays by an order of magnitude, $2(Z_{\rm Zn}-Z_{\rm Ga})/(Z_{\rm Zn}+Z_{\rm Gz})=3.3$\%. 

   Fig.~\ref{fig:diffraction}(b) shows the low-temperature neutron powder diffraction data for \YbZnGaO. A high-quality Rietveld refinement is achieved within the reported space group $P6_3/mmc$ with \SI{1.14}{wt\%} $\rm Yb_5Ga_3O_{12}$ and \SI{0.47}{wt\%} $\rm Yb_2O_3$ as secondary phases. Allowing Zn$^{2+}$ and Ga$^{3+}$ to partially occupy each other's Wyckoff sites  while enforcing charge neutrality with the following stoichiometry: $\rm Yb(Zn_{1-y}Ga_y)_2(Ga_{1-x}Zn_x)O_{5+y-x/2}$, yields a best fit with significant site mixing $(x,y)=(0.60(1),0.35(1))$, as shown in Fig.\ref{fig:neutron_diffraction_chi2}. The refinement parameters are listed in Table ~\ref{tab:neutron_diffraction}. Notably, a comparable fit was obtained without enforcing charge neutrality, allowing the occupancies of all oxygen sites to vary freely. In this case, the optimal fit was found at $(x,y)=(0.55,0.36)$. The near random Zn$^{2+}$/Ga$^{3+}$ site occupancy is not surprising given the similar sizes of these ions and the evidence for such site mixing in oxides and intermetallics including $\rm YbZnGaO_4$\cite{PhysRevLett.120.087201}, $\rm ZnGa_2Se_4$\cite{HANADA1997373}, and $\rm GdZn_2Ga_2$\cite{VERBOVYTSKYY2012106}.

    \begin{table}[t]
    \begin{ruledtabular}
        \begin{tabular}{ccccccc}
            \multicolumn{7}{c}{\YbZnGaO\ Neutron Powder Diffraction (T=12K)} \\
            \hline
            \multicolumn{2}{c}{Space group}                                  & \multicolumn{5}{|c}{\textit{P6$_3$/mmc}}                                                         \\
            \hline
            \multicolumn{2}{c}{\multirow{2}{8em}{Cell parameters}}           & \multicolumn{5}{|c}{$a=b=3.37360(6)$~\AA, $c=21.9271(4)$~\AA }                             \\
            \multicolumn{2}{c}{}                                             & \multicolumn{5}{|c}{$\alpha = \beta =$ \SI{90}{\degree}, $\gamma =$ \SI{120}{\degree}} \\
            \hline
            \multicolumn{2}{c}{Fit quality}                                  & \multicolumn{5}{|c}{$R_{p}=$ \num{6.61}, $R_{wp}=$ \num{5.99}, $\chi^{2}=$ \num{5.10}}    \\
            \hline
            Atom  & x & y & z & Occ. & $U_{11}$(\AA$^2$) & $U_{33}$(\AA$^2$)\\
            \hline
            Yb1    & 0                  & 0           & 0            & 1.00& 0.0009(2) &0.0215(2) \\
            Ga1   & 0                  & 0           & 1/4            & 0.40(1) & 0.0011(3) & 0.0029(3) \\

            Zn2    & 0                  & 0           & 1/4            & 0.60(1) & 0.0011(3) & 0.0029(3)\\
            
            Zn1  & 1/3                  & 2/3           & 0.1376(2)            & 0.65(1) & 0.0020(2) & 0.0100(3)\\
            Ga2       & 1/3                  & 2/3           & 0.1376(2)            & 0.35(1) & 0.0020(2) & 0.0100(3)\\
            
            O1    & 1/3                  & 2/3           & 0.0487(3)            & 1.007(3) & 0.0057(4) & 0.0089(3) \\
            O2   & 0                  & 0           & 0.1546(2)            & 0.999(2) & 0.0095(4) & 0.0132(3) \\
            O3      & 1/3                  & 2/3           & 1/4           & 1.045(4) & 0.0074(6) & 0.0576(8) \\
        \end{tabular}
        \end{ruledtabular}

        \caption{The results of Rietveld refinement of neutron  diffraction from a powder sample of \YbZnGaO\ at $T=$ \SI{12}{K}. Site mixing was refined in accordance with the chemical formula $\rm Yb(Zn_{1-y}Ga_y)_2(Ga_{1-x}Zn_x)O_{5+y-x/2}$ that constrains the oxygen stoichiometry to ensure charge neutrality.}
        \label{tab:neutron_diffraction}
    \end{table}

    \begin{figure}[t]
        \centering
        \includegraphics[width=0.9\linewidth]{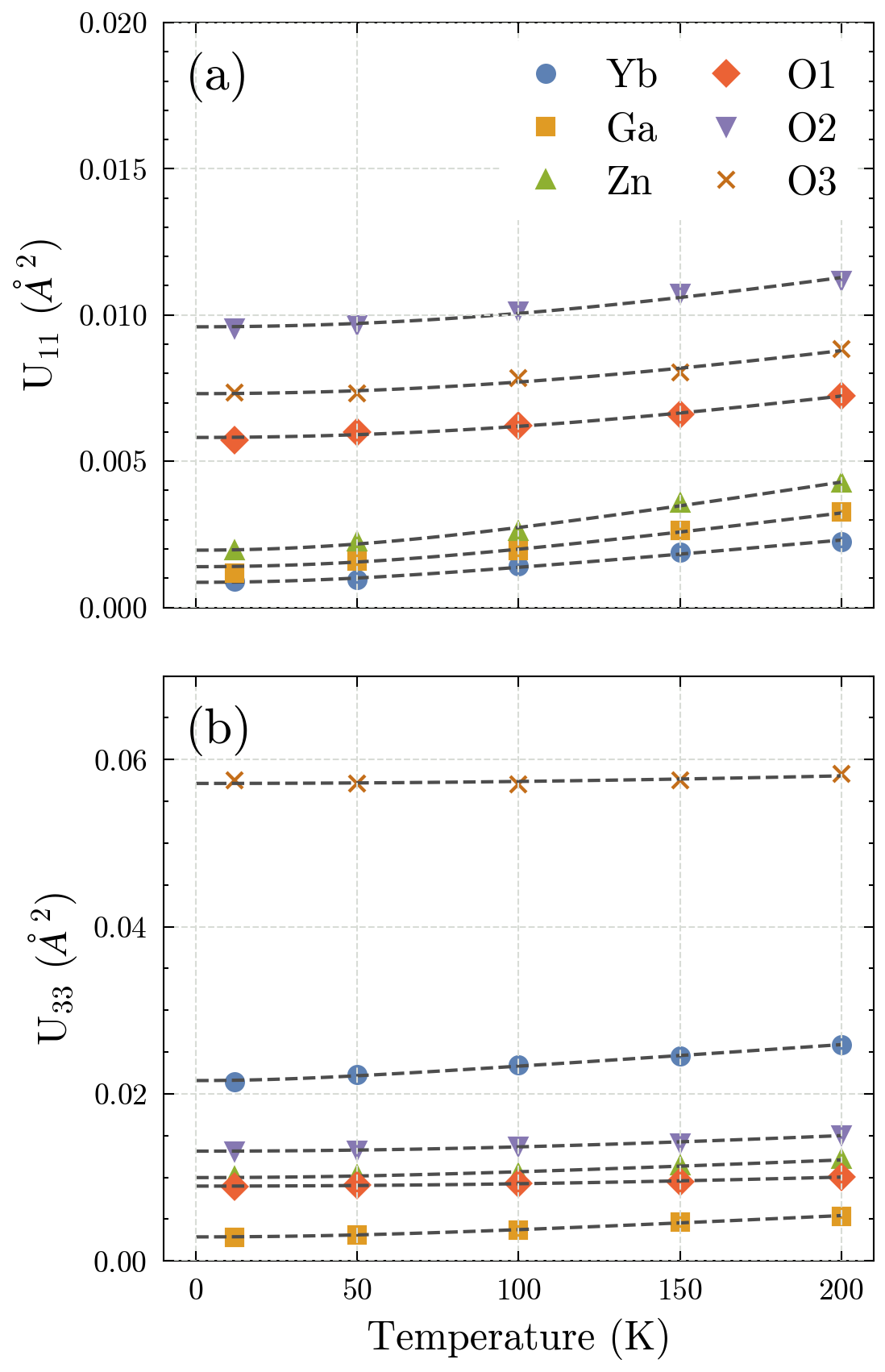}
        \caption{Temperature dependence of the mean-squared atomic displacement in the crystallographic (a) $a$ direction ($U_{11}$), and (b) $c$ direction ($U_{33}$), of all sites in \YbZnGaO, obtained from Rietveld refinement of neutron powder diffraction data. Black dashed lines represent fits to a model combining the Debye model for thermal vibrations and a temperature-independent static component (Eq.\ref{eq:Uiso}). The fitted Debye temperatures and static displacements for each site are shown in Fig.\ref{fig:Debye_temp_all_atoms}. }
        \label{fig:U_all_atoms}
    \end{figure}

    \begin{figure}[t]
        \centering
        \includegraphics[width=0.9\linewidth]{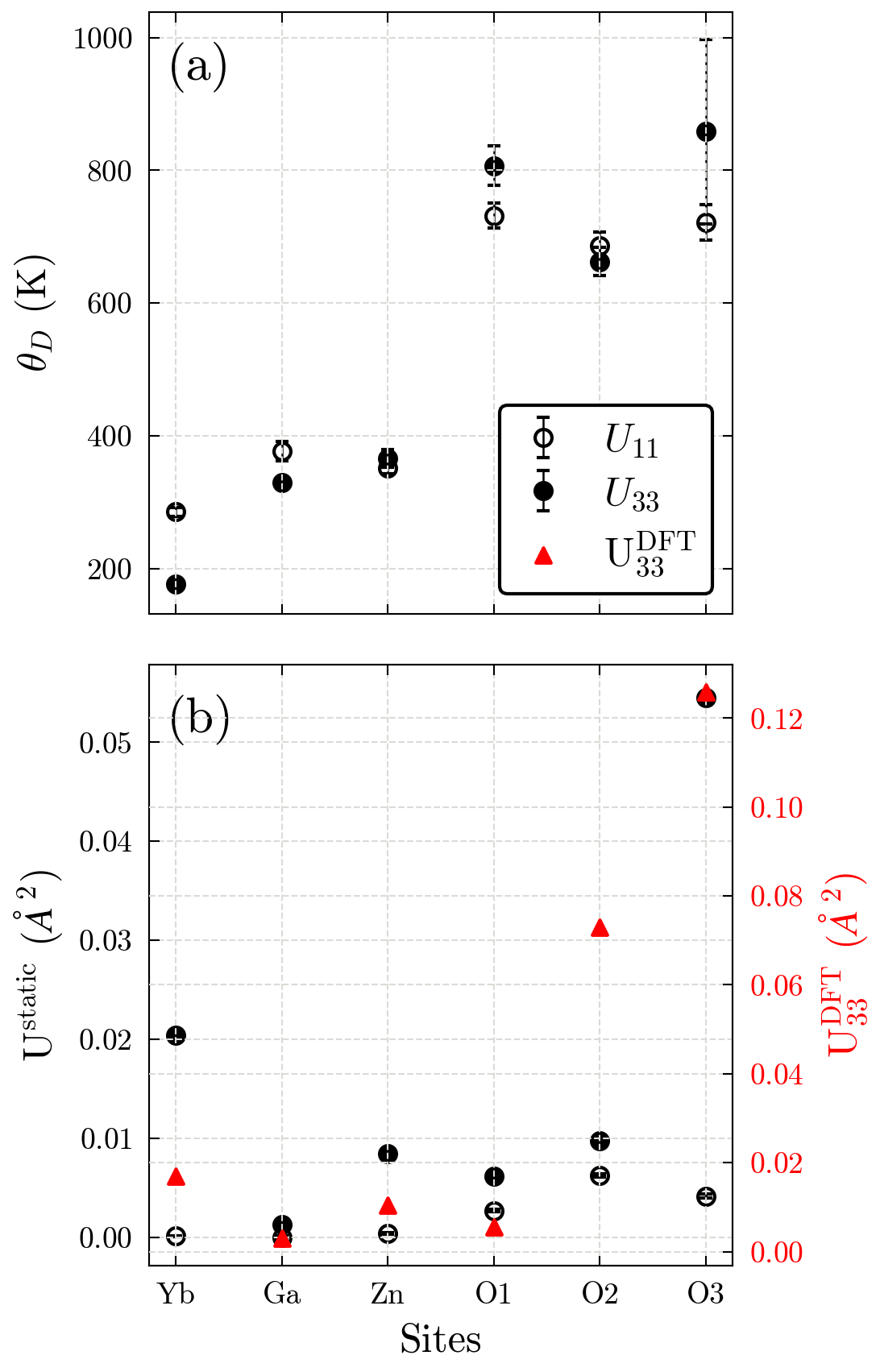}
        \caption{(a) Debye temperatures $\rm \theta_D$ and (b) static contributions to the mean-squared atomic displacement, $\rm U^{static}$, for all atomic sites in \YbZnGaO, fitted from temperature dependent neutron powder diffraction data similar to Fig.~\ref{fig:diffraction}(b). Open and filled circles represent values along the crystallographic $a$ ($ U_{11}$) and $c$ ($ U_{33}$) directions, respectively. Red triangles are $\rm U_{33}$ calculated with density functional theory by considering the random distribution of $\rm Ga^{3+}/Zn^{2+}$ ions across the two Wyckoff sites. The large static displacement of the Yb site along the $c$ axis indicates distortions of the YbO$_6$ octahedra, which leads to shifts in the crystal electric field(CEF) excitation energies. These local distortions also generate variations in Yb-O bond lengths and Yb-O-Yb bond angles, which will lead to quenched randomness in the magnetic exchange interactions between \Yb\ ions. Site dependence of the Land\'{e} $g$-factor for Yb$^{3+}$ is also anticipated. This will produce heterogeneous broadening of Zeeman split excitations in high magnetic fields.}
        \label{fig:Debye_temp_all_atoms}
    \end{figure}

    \begin{figure}[t]
        \centering
        \includegraphics[width=0.9\linewidth]{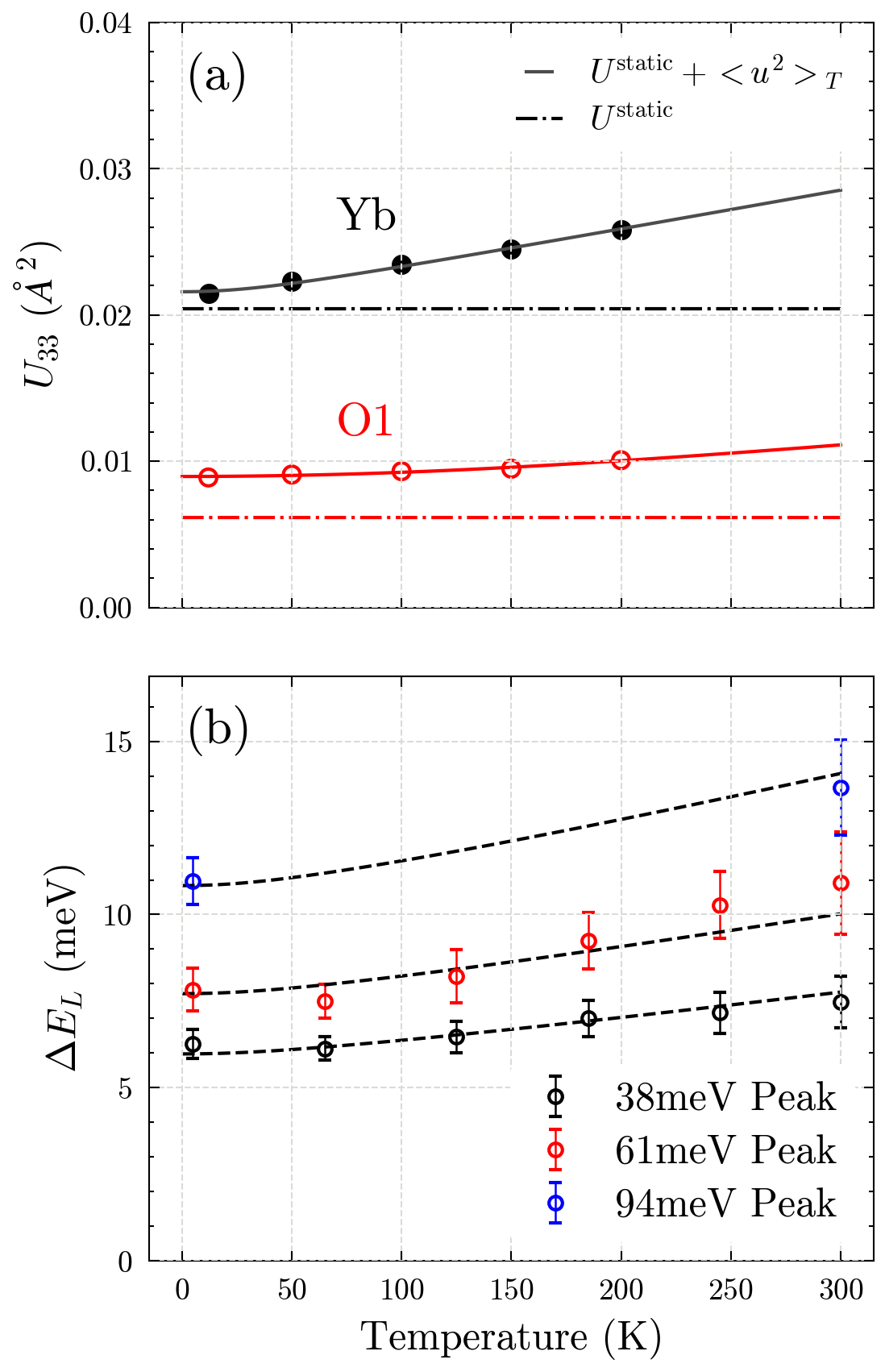}
        \caption{(a) Temperature dependence of the mean-squared atomic displacement of \Yb\ and its coordinating \oxygen\ along  the crystallographic $c$ direction, $U_{33}^{Yb}$ and $U_{33}^{O}$, obtained from Rietveld refinement of neutron powder diffraction data. The same data are compared to those for other atoms and for in-plane displacements in Fig.\ref{fig:U_all_atoms}. Solid lines are fits to a model combining the Debye model for thermal vibrations and a temperature-independent static distribution $U^{\rm static}$. Debye temperatures of $\theta_D^{Yb}=$ \SI{177}{K} was found for Yb and $\theta_D^{O}=$ \SI{807}{K} for O. Horizontal dashed lines correspond to the fitted $U_{\rm static}^{Yb}=$ \SI{0.02}{\angstrom ^2} for Yb, and $U_{\rm static}^{O}=$ \SI{0.006}{\angstrom ^2} for O. These static displacements dominate over quantum fluctuations in the $T \rightarrow 0$ limit. (b) Temperature dependence of $\Delta E_{\rm L}$, the Lorentzian component of the FWHM of each  CEF transition. Dashed lines are the sum of the fits in (a) scaled with an overall factor $A$. The agreement between the temperature dependence of $\Delta E_{\rm L}$ and $U_{33}$ indicates that the physical broadening of the CEF peaks is due to the combination of static and dynamic  displacements  of \Yb\ and its coordinating \oxygen\ atoms along $\bf c$.}
        \label{fig:U33_FWHM_vs_T}
    \end{figure}

\section{Analysis \& Discussion}

\subsection{Crystal Field Levels}
    In \YbZnGaO, the crystal electric field (CEF) lifts the degeneracy of the $J = 7/2$ multiplet of \Yb, splitting it into four Kramers doublets. The CEF Hamiltonian can be expressed in the most general form as
        \begin{equation}
        H_{\rm CEF}=\sum_{n,m} B^m_n O_n^{m},
        \label{eq:H_CEF}
        \end{equation}
    where $O_n^{m}$ are the Stevens operators and $B^m_n$ are adjustable scalar CEF parameters encapsulating the strength and symmetry of the interaction~\cite{stevens_1952}. The $D_{3d}$ symmetry of the Yb site gives six symmetry-allowed non-zero CEF parameters: $B_2^0,B_4^0,B_4^3,B_6^0,B_6^3$, and $B_6^6$. The ground-state doublet can be treated as an effective spin-1/2 degree of freedom at low temperatures due to the large energy gap (\SI{38}{meV}) from the excited states. The nature of the ground-state doublet sets the magnetic anisotropy and $g$-tensor of the effective spin, directly influencing the low-energy spin dynamics and exchange interactions. Therefore, understanding the CEF Hamiltonian and its eigenstates is essential for interpreting the magnetic excitations and potential quantum spin liquid behavior in this material.
    
    To determine the CEF level scheme of Yb$^{3+}$ in \YbZnGaO, we fitted the inelastic neutron scattering spectrum shown in Fig.~\ref{fig:INS_Ecut} with $H_{\rm CEF}$. We constrained the fit with previously reported anisotropic Land\'e $g$-factors of $g_{\parallel}=$ \SI{3.44}{} and $g_{\perp}=$ \SI{3.04}{} obtained from saturation magnetization measurements\cite{PhysRevLett.133.266703}, ensuring the number of independent observable quantities (7) exceeds the number of fitted $B^m_n$ parameters (6) and resulting in a fully constrained fit. We note that the $B^m_n$ parameters cannot be uniquely determined without the inclusion of the Land\'e $g$-factor constrains, as expected from the limited number of independent observables relative to the number of free parameters in the unconstrained fit and previous reports on other \Yb\ systems\cite{10.21468/SciPostPhysCore.5.1.018}.

    A unique best fit was obtained and the fitted $B^m_n$ are listed in Table \ref{tab:Bmn}. Uncertainties on each $B^m_n$ were determined by allowing variations that yield a reduced $\chi^2$ no greater than $(1+\frac{1}{N-n})\chi^2_{\rm min}$, where $N$ is the number of data point in the INS spectrum, $n$ is the number of free parameters in the fit, and $\chi^2_{\rm min}$ is the reduced $\chi^2$ value at the best fit. During this procedure, all other fit parameters were allowed to relax freely to their locally optimal values. The resulting eigenvalues and eigenvectors of the fitted $H_{\rm CEF}$ are provided in Table~\ref{tab:Eigenvectors}, with uncertainties propagated from the fitted $B^m_n$ ranges.

    \begin{table}[b]
        \begin{tabular}{c|cc}
        \hline
        \hline
            $B^m_n$ & Fit values (meV)& Point charge model (meV)   \\
            \hline
             $B^2_0$& \SI{-0.91(3)}{} & \SI{0.47}{}\\
             $B^4_0$& \SI{1.46(2)e-2}{} & \SI{2.35e-2}{}\\
             $B^4_3$&  \SI{-7.5(2)e-1}{} &  \SI{-9.2e-1}{}\\
             $B^6_0$& \SI{6.0(3)e-4}{} & \SI{3.7e-4}{} \\
             $B^6_3$&  \SI{-3.1(5)e-2}{}&  \SI{-2.13e-3}{}\\
             $B^6_6$&  \SI{1.83(6)e-2}{}&  \SI{3.66e-3}{}\\
         \hline
        \hline
        \end{tabular}
    \caption{(Left column) Crystal electric field(CEF) Hamiltonian  parameters for the \Yb\ ion in \YbZnGaO\ fitted from the inelastic neutron scattering spectrum averaged over $Q \in [4,5]$ \SI{}{\angstrom^{-1}} (Fig. ~\ref{fig:INS_Ecut}) and Land\'e $g$-factors obtained from saturation magnetization measurements. (Right column) CEF Hamiltonian parameters obtained from the point charge model by considering the 9 nearest atomic neighbors.}
    \label{tab:Bmn}
    \end{table}

    \begin{table}[t]
            \begin{ruledtabular}
            \begin{tabular}{cc}
                Eigenvalues & \multirow{2}{4em}{Eigenvectors}  \\
                (meV) & \\
                \hline
                0.0 & $0.47(2)| \pm\frac{1}{2}\rangle \mp 0.47(3)| \mp\frac{5}{2}\rangle \pm 0.75(2)| \pm\frac{7}{2}\rangle  $ \\
                38(3) & $0.37(3)| \pm\frac{1}{2}\rangle \pm0.87(2)| \mp\frac{5}{2}\rangle \pm0.32(2)| \pm\frac{7}{2}\rangle$ \\
                60(3) & $|\pm\frac{3}{2}\rangle$ \\
                94(4) & $ \pm 0.80(1)|\pm\frac{1}{2}\rangle-0.13(3)|\mp\frac{5}{2}\rangle  -0.58(2)| \pm\frac{7}{2}\rangle$\\
            \end{tabular}
            \end{ruledtabular}
            \caption{Eigenvectors and eigenvalues of the fitted crystal field Hamiltonian $H_{\rm CEF}$ of the \Yb\ ion in \YbZnGaO. Uncertainties are propagated from the fitted $B^m_n$ parameters.}
            \label{tab:Eigenvectors}
        \end{table}

\subsection{Static and Dynamic Anisotropic Atomic Displacement}

    Zn$^{2+}$/Ga$^{3+}$ site mixing introduces quenched charge disorder, which is expected to distort the oxygen coordination environment around \Yb\ and result in heterogeneous broadening of the CEF levels. To quantify this distortion, we examine the anisotropic atomic displacement parameters $U_{ij}$, obtained through refinement of the Debye-Waller factors from neutron diffraction data, given by $ \exp(-\mathbf{Q}^\intercal \mathbf{U} \mathbf{Q})$. The temperature-dependent displacement parameters $U_{11} = U_{22}$ and $U_{33}$ in Cartesian coordinates are shown for all atomic sites in Fig.~\ref{fig:U_all_atoms}. We observe that displacements along the $c$-axis are significantly larger than those within the basal plane, and even in the low-temperature limit ($T \rightarrow 0$), there is a broad range of values across the three oxygen Wyckoff sites. This strongly indicates the presence of static structural disorder originating from $\rm Ga^{3+}/Zn^{2+}$ site mixing.

     We distinguish two contributions to the anisotropic atomic displacement
    $U_{ij}$:
    \begin{equation}
        U_{ij}=  U^{\rm static}_{ij}+ <u^{2}>_{T}, \label{eq:Uiso}
    \end{equation}
    where $ U^{\rm static}$ represents the static displacement of the atoms due to quenched structural disorder (site mixing) and $<u^{2}>_{T}$ is the vibrational(phononic) component along one Cartesian direction. Under the harmonic approximation,
    \begin{equation}
        <u^{2}>_{T}= \frac{\hbar}{2M}\int_{0}^{\infty}\rho(\omega)\text{coth}(\frac{\hbar
        \omega}{2k_{B}T}) \frac{d\omega}{\omega},
    \end{equation}
    where $M$ is the atomic mass and $\rho(\omega)$ is the phonon density of states. We use the Debye approximation to the phonon density of states as $\rho (\omega)=3\omega^{2}/\omega_{D}^{3} \Theta(\omega_D-\omega)$, where $\Theta(\omega)$ is the Heaviside function and $\omega_{D}$ is the Debye frequency. While this form is designed to approximate the acoustic phonon spectrum, we adopt it as an approximation to the full spectrum of vibrations for each Wyckoff site. We fit  $ U_{11}(T)$ and $ U_{33}(T)$ obtained from diffraction for each Wyckoff site to Eq.~\ref{eq:Uiso}, and show the resulting site-specific  Debye temperatures $\theta_D=\hbar\omega_D/k_B$ and static displacements $U^{\rm static}$ in Fig.\ref{fig:Debye_temp_all_atoms}. As anticipated on general grounds, the Debye temperatures scale with $1/\sqrt{M}$. Specifically, averaging over $U_{11}$ and $U_{33}$ and all Wyckoff sites, we obtain $\langle\omega^2_D M\rangle=~$\SI{250(60)}{N/m}, where the bracket indicates the standard deviation over the ensemble. 

    \subsection{Correlation Between CEF Broadening and Structural Disorder}

    The significant contribution of $U^{\rm  static}$ to $U_{ij}$ indicates the presence of quenched disorder such as Zn$^{2+}$/Ga$^{3+}$ site mixing. The displacements of \Yb\ and its coordinating \oxygen\ ions should lead to heterogeneity in the CEF levels. To determine whether this is consistent with the observed temperature-dependent broadening of the CEF transitions (Fig.~\ref{fig:INS_Ecut}), we first extract a quantitative measure of the physical width $\Delta E_{\rm L}$ by fitting the CEF peaks with a Voigt function with the Gaussian FWHM fixed at the resolution width $\Delta E_{\rm res}$. Fig.~\ref{fig:U33_FWHM_vs_T} compares the temperature dependence of $\Delta E_{\rm L}$ with that of $\rm U_{33}$ for the \Yb\ and its coordinating \oxygen\ ions. The dashed line fit in Fig.~\ref{fig:U33_FWHM_vs_T}(b) shows that $\Delta E_{\rm L}$ follows the same temperature dependence as $U_{33}$. Specifically,  
    $\Delta E_{{\rm L},n}(T) = A_nE_{{\rm exp},n}(U_{33}^{Yb}(T)+U_{33}^{O}(T))$, with $A_n=$~\SI{5.2}{\angstrom^{-2}}, \SI{4.2}{\angstrom^{-2}}, and \SI{3.8}{\angstrom^{-2}} for CEF excitations $n=1,2,3$, respectively. The excellent fit and the similarity of $A_n$ suggests that the broadening of the CEF peaks results from a combination of static and dynamic displacement of \Yb\ and its surrounding \oxygen\ ions  along $\bf c$. The static disorder-related part of $\Delta E_{{\rm L},n}$, denoted as $\Delta E_{\text{static},n}=A_n U_{{\rm static},n}$, is listed in Table \ref{tab:CEF_FWHM}.

    \begin{table*}[]
        \begin{ruledtabular}
        \begin{tabular}{cccccccc}
            Crystal field level & $E_{\rm exp}$ (meV) & $\Delta E_{\rm exp}$ (meV) & $\Delta E_{\rm res}$ (meV) &$\Delta E_{\rm L}$ (meV)& $\Delta E_\text{static}$ (meV) & $E_{\rm PCM}$ (meV)&$\Delta E_{\rm PCM}$ (meV) \\
            \hline
            1                   & 38.0(1)             & 7.5(3)                   & 2.4  &6.7(4)                 & 5.0                 & 37 & 5.0                    \\
            \hline
            2                   & 60.5(2)             & 9.4(5)                   & 2.2  &8.8(5)                 & 6.5                 & 45&8.0                    \\
            \hline
            3                   & 94.3(2)             & 10.1(4)                  & 2.0  &9.7(4)                 & 9.1                 & 91&8.8                    \\
        \end{tabular}
        \end{ruledtabular}
        \caption{Energy and full width at half maximum (FWHM) of crystal field excitation peaks in \YbZnGaO. $\Delta E_{\rm exp}$ is extracted using a fit of the $T=5$~K inelastic neutron scattering data in Fig.~\ref{fig:INS_Ecut} to the scattering cross section associated with the Yb$^{3+}$ crystal field Hamiltonian. Each delta-function was replaced by a normalized Voigt line shape with a transition-specific Lorentzian FWHM $\Delta E_{\rm L}$ and the  instrumental FWHM $\Delta E_{\rm res}$ listed in the table as the Gaussian FWHM. $\Delta E_\text{static}$ is the static component of $\Delta E_{\rm L}$, extracted from a fit to the temperature-dependent data in Fig.~\ref{fig:U33_FWHM_vs_T} as described in the text. $E_{\rm PCM}$ is the eigen-energies of the point-charge Hamiltonian of \Yb\ ions. $\Delta E_{\rm PCM}$ is the static broadening calculated with the point charge model by considering the random static displacement of \Yb\ and its $\rm O^{2-}$ ligands from their nominal positions.}
        \label{tab:CEF_FWHM}
    \end{table*}

  To determine whether the observed static structural disorder can account for the magnitude of the CEF level broadening in \YbZnGaO, we simulated the impact of local distortions on the CEF spectrum by introducing random perturbations to the $z$-coordinates of the \Yb\ and its coordinating \oxygen\ ions. The perturbations were drawn from a Gaussian probability distribution with a standard deviation of  $\sqrt{U_{33}}$. The resulting distorted geometries were used to construct CEF Hamiltonians using a point-charge model, implemented with the \textit{PyCrystalfield} package\cite{Scheie2021,hutchings_1964}. Although the point-charge description of the CEF Hamiltonian neglects effects such as the spatial distribution of charges on the ions, covalency of ligand bonding, and the screening of the magnetic electrons by the outer electron shells, it provides a quantitative framework to relate atomic scale structure with the CEF spectrum. We include the nine nearest atomic neighbors in the calculation of the CEF levels and Table \ref{tab:CEF_FWHM} shows that this simplified model provides semi-quantitative agreement with the measured level scheme. The point charge model CEF Hamiltonian parameters are listed in Table \ref{tab:Bmn}. 

    The randomly perturbed \Yb\ and $\rm O^{2-}$ ions yielded a distribution of YbO$_{6}$ octahedron geometries, each producing  three CEF excitations at slightly different energies, mimicking the heterogeneous distribution of level schemes that can be anticipated in a sample with the displacements inferred from diffraction. The simulated FWHM values of the three CEF peaks due to this effect, $\Delta E_\text{PCM}=$ \SI{5.0}{meV}, \SI{8.0}{meV}, and \SI{8.8}{meV}, respectively, are listed in Table \ref{tab:CEF_FWHM} and compare favorably with the physical widths inferred from inelastic magnetic neutron scattering. For a more accurate picture of the local structure, atomic pair density function analysis of diffraction data can be employed\cite{doi:10.1126/science.1135080}, potentially allowing for a more exact modeling of disorder effects on the CEF level scheme. 

    We have proposed that the static displacements of  \Yb\ and \oxygen\ ions are a result of the charge disorder created by  Zn$^{2+}$/Ga$^{3+}$ site mixing. To challenge this hypothesis, we constructed 11 distinct unit cells of \YbZnGaO, realizing all possible $\rm Ga^{3+}/Zn^{2+}$ distributions across the two Wyckoff sites that preserve the overall stoichiometry. The atomic positions were then relaxed from the nominal and average positions through ab-initio Density Functional Theory as described in Section~\ref{subsec:DFT}. The relaxed positions indeed vary across the Zn$^{2+}$/Ga$^{3+}$ configurations and from this we obtain mean-squared displacements for each Wyckoff site relative to the nominal position. Reported as red triangular symbols in Fig.~\ref{fig:Debye_temp_all_atoms}, we find that DFT overestimates the displacements by as much as a factor 2. We speculate that this may be because the imposed unit cell symmetry does not allow for the degree of screening nor the full range of possible Zn$^{2+}$/Ga$^{3+}$ configurations to adequately model the real structure. In addition, the modeling approach we used does not incorporate correlations in the distribution of Zn$^{2+}$/Ga$^{3+}$, which would tend to screen charge and reduce the resulting displacements. Yet the DFT modeling does account for the pronounced displacement of the Yb and O3 sites and the reduced displacement for Zn and Ga sites, which is consistent with experimental observations and therefore supports the hypothesis that the disorder is driven by Zn$^{2+}$/Ga$^{3+}$ site mixing.

    \section{Conclusion}

    In summary, the crystal electric field excitations of \Yb\ in the quantum spin liquid candidate \YbZnGaO\ have been studied with inelastic neutron scattering in powder samples grown by conventional solid-state synthesis. All three CEF transitions exhibit pronounced physical broadening. By combining temperature-dependent inelastic neutron scattering with neutron powder diffraction, we identified a substantial static disorder-related contribution to this broadening. We ascribe this to a disordered local charge environment around the \Yb\ ions induced by chemical site mixing between Zn$^{2+}$ and Ga$^{3+}$ ions. We supported this explanation with neutron diffraction, which shows pronounced $\rm Ga^{3+}/Zn^{2+}$ site mixing and anisotropic static displacement of Yb and its coordinating O ions from their nominal positions. We have not explored to what extent synthesis conditions and/or annealing protocols impact $\rm Ga^{3+}/Zn^{2+}$ site mixing. However, we note that the widths of the crystal field excitations in our sample, which we have shown are closely linked to site mixing, appear to be  similar to those reported for powder samples in the paper that inspired this project\cite{PhysRevLett.133.266703}.
    
    We connected our experimental evidence of structural disorder with the linewidth of CEF excitations through the point-charge model of the crystal field Hamiltonian for \Yb. In particular, the observed displacements of Yb and its coordinating oxygen ligands accounts for the CEF linewidths at a semi-quantitative level. First-principle DFT calculations furthermore indicates that the random distribution of Zn$^{2+}$ and Ga$^{3+}$ ions leads to the observed displacements of \Yb\ and \oxygen\ ions.

    These displacements change the Yb-O bond lengths and Yb-O-Yb bond angles and therefore also modify the magnetic super-exchange interactions. A zeroth order model to describe the magnetism of \YbZnGaO\ might then be a triangular lattice of spin-1/2 degrees of freedom with nearest-neighbor ($J_1$) and next-nearest-neighbor ($J_2$) antiferromagnetic interactions that are both subject to quenched randomness. Randomness in the Land\'{e} $g$-factor, which influences high field properties, can also be anticipated. 
    
    A recent theoretical work\cite{PhysRevB.99.085141} studied this model using exact diagonalization and density-matrix renormalization group (DMRG) techniques with a single dimensionless parameter $\Delta$ controlling the variance of the interactions. It was found that increasing $\Delta$ expands the range of $J_2/J_1$ where $T=0$ magnetic order is suppressed producing a spin-liquid-like phase without spin-glass freezing. 
    
    Our work indicates that \YbZnGaO\ should be considered within this larger phase diagram of quantum magnetism that includes bond disorder $\Delta$. If methods can be devised to vary the non-magnetic cation site mixing in \YbZnGaO\ or related compounds, it would be of great interest to explore the $\Delta$-dependence of entanglement and emergent quasi-particle dynamics near the QSL regime through neutron scattering and other experimental techniques even if the $\Delta\rightarrow 0$ limit cannot be accessed.

    \begin{acknowledgments}
   We gratefully acknowledge stimulating discussions with Sara Haravifard and Pengcheng Dai. This work was supported by the Department of Energy, Office of Science, Basic Energy Sciences under Award No. DE-SC0024469. C.B. was supported by the Gordon and Betty Moore Foundation EPIQS program under GBMF9456. The Institute for Quantum Matter was supported through a generous donation by William H. Miller III to the Department of Physics and Astronomy. This research used resources at the Spallation Neutron Source, a DOE Office of Science User Facility operated by Oak Ridge National Laboratory. The beam time was allocated to POWGEN and SEQUOIA through proposal number IPTS-33395.1. Calculations were performed using computational resources of the Maryland Advanced Research Computing Center and the Advanced Research Computing at Hopkins (ARCH) Rockfish cluster. The Rockfish cluster is supported by the National Science Foundation Award No. OAC-1920103. 
    \end{acknowledgments}

    \bibliography{bib}
\end{document}